\documentclass[proceedings]{JHEP} % 10pt is ignored!

\usepackage{epsfig}                   % please use epsfig for figures
\newcommand{\be}{\begin{equation}}
\newcommand{\ee}{\end{equation}\noindent}
\newcommand{\bear}{\begin{eqnarray}}
\newcommand{\ear}{\end{eqnarray}\noindent}
\newcommand{\no}{\noindent}
\date{}

\newcommand{\slD}{\raise.15ex\hbox{$/$}\kern-.57em\hbox{$D$}}
\newcommand{\slpartial}{\raise.15ex\hbox{$/$}\kern-.57em\hbox{$\partial$}}
\newcommand{\slG}{{{\dot G}\!\!\!\! \raise.15ex\hbox {/}}}

\def\GBd12{{\dot G}_{B12}}

\def\non{\nonumber}
\def\beqn*{\begin{eqnarray*}}
\def\eqn*{\end{eqnarray*}}

\def\square{\kern1pt\vbox{\hrule height 1.2pt\hbox{\vrule width 1.2pt
   \hskip 3pt\vbox{\vskip 6pt}\hskip 3pt\vrule width 0.6pt}
   \hrule height 0.6pt}\kern1pt}
\def\slash#1{#1\!\!\!\raise.15ex\hbox {/}}

\def\dps{\displaystyle}

\def\half{{1\over 2}}

\def\fourth{{1\over4}}

\def\FFdual{F\cdot\tilde F}

\def\e{\mbox{e}}

\def\kinb{{1\over 4}\dot x^2}

\def\PITD{{(4\pi T)}^{-{D\over 2}}}
\def\4piTD{{(4\pi T)}^{-{D\over 2}}}
\def\4piT4{{(4\pi T)}^{-2}}

\def\Tintm4{{\dps\int_{0}^{\infty}}{dT\over T}\,e^{-m^2T}
    {(4\pi T)}^{-2}}
\def\Tintm{{\dps\int_{0}^{\infty}}{dT\over T}\,e^{-m^2T}}
\def\Tint{{\dps\int_{0}^{\infty}}{dT\over T}}

\def\Dx{\dps\int{\cal D}x}

\def\Dpsi{\dps\int{\cal D}\psi}
\def\Tr{{\rm Tr}\,}

\def\FFdual{F\cdot\tilde F}
\def\bbbz{{\mathchoice {\hbox{$\sf\textstyle Z\kern-0.4em Z$}}
{\hbox{$\sf\textstyle Z\kern-0.4em Z$}}
{\hbox{$\sf\scriptstyle Z\kern-0.3em Z$}}
{\hbox{$\sf\scriptscriptstyle Z\kern-0.2em Z$}}}}
\conference{Corfu Summer Institute on Elementary Particle Physics,
1998\hspace{20pt}
}

\title{Axial Vector Amplitudes, Second Order Fermions, and
Standard Model Photon -- Neutrino Processes}

\author{Christian Schubert 
\\
        Laboratoire d' Annecy-le-Vieux de Physique Th{\'e}orique LAPTH,
  Chemin de Bellevue, BP 110, F-74941 Annecy-le-Vieux CEDEX, France\\
        E-mail: \email{schubert@lapp.in2p3.fr}}

\abstract{We present an extension of the
string-inspired technique suitable to the calculation
          of amplitudes and effective Lagrangians involving both
          vector and axial vector gauge fields. The technique is easily
          adaptable to problems involving constant external fields.
          We demonstrate the advantages of the formalism by a
          calculation of the one-loop vector-axialvector amplitude
          in a general constant electromagnetic field. The relevance
          of this calculation for photon-neutrino processes is
          commented upon. We also clarify the properties of the 
          formalism with respect to the chiral symmetry,
          and its connection to second order fermions.\\
$\phantom{xxxxxxxxxxxxxxxxxxxxxxxxxxxxxxxxxxxxxxxxxxxxxxxxxxxxxxxx} {\rm LAPTH-Conf-729/99}$
}

\begin{document}

  \section{Introduction}

The idea of using string theory methods for the
calculation of amplitudes in ordinary field theory
was originally introduced in the
context of QCD \cite{grscbr,berkos} and quantum 
gravity \cite{bedush}. In particular, it was used for
the first complete calculations of the 
one-loop five gluon amplitude \cite{5glu}
and the one-loop four graviton amplitude \cite{dunnor}. 
Later it was found that even in QED similar techniques 
can lead to
significant improvements over standard field theory
methods \cite{strassler,ss13,dashsu}, particularly
for processes involving constant external fields
\cite{mckshe,cadhdu,shaisultanov,rss,adlsch,frss,korsch}.

The present talk is devoted to an extension of the
``string-inspired'' technique suitable
to the calculation of processes involving both
vector and axial-vector couplings \cite{mcksch,dimcsc}.
In contrast to previous attempts at such a generalization
\cite{mnss,dhogag} this extension allows one to
treat the real and the imaginary part of these
amplitudes in a unified way.

In the introduction we first give some physical motivation for the
study of vector -- axialvector amplitudes. 
In section 2 we explain the second-order formalism for
spinor QED \cite{feygel,hostler,morgan} and its recent extension
to the vector-axial\-vec\-tor case \cite{mcksch,dimcsc}.
The second-order formalism is then used in section 3 to
derive a path-integral representation for the one-loop
effective action due to a Dirac fermion loop coupled
to abelian vector and axialvector backgrounds. 
We explain in detail how this path integral can be used
for the calculation of one-loop $N$ - point amplitudes.
In section 4
we discuss the behaviour of the resulting formalism 
with respect to 
the ``$\gamma_5$ - problem'' of dimensional regularisation,
including a recalculation of the ABJ anomaly.
In Section 5 we apply this technique to
calculations in constant background fields.
We give explicit results for the 
Scalar and Spinor QED vacuum polarisation
tensors in a general constant field, as well as for the
corresponding vector-axialvector amplitude.
Our conclusions are given in section 6.

\begin{figure}[ht]
\begin{center}
\begin{picture}(0,0)%
\epsfig{file=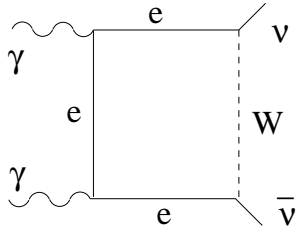}%
\end{picture}%
\setlength{\unitlength}{0.00047500in}%
\begingroup\makeatletter\ifx\SetFigFont\undefined
% extract first six characters in \fmtname
\def\x#1#2#3#4#5#6#7\relax{\def\x{#1#2#3#4#5#6}}%
\expandafter\x\fmtname xxxxxx\relax \def\y{splain}%
\ifx\x\y   % LaTeX or SliTeX?
\gdef\SetFigFont#1#2#3{%
  \ifnum #1<17\tiny\else \ifnum #1<20\small\else
  \ifnum #1<24\normalsize\else \ifnum #1<29\large\else
  \ifnum #1<34\Large\else \ifnum #1<41\LARGE\else
     \huge\fi\fi\fi\fi\fi\fi
  \csname #3\endcsname}%
\else
\gdef\SetFigFont#1#2#3{\begingroup
  \count@#1\relax \ifnum 25<\count@\count@25\fi
  \def\x{\endgroup\@setsize\SetFigFont{#2pt}}%
  \expandafter\x
    \csname \romannumeral\the\count@ pt\expandafter\endcsname
    \csname @\romannumeral\the\count@ pt\endcsname
  \csname #3\endcsname}%
\fi
\fi\endgroup
\begin{picture}(3834,2736)(120,-1050)
\end{picture}
\caption{\label{ggnnbar}
Diagram contributing to $\gamma\gamma\to\nu\bar\nu$
.}
\end{center}
\end{figure}

Processes involving photons and neutrinos are presently
of importance primarily for astrophysics and
cosmology \cite{raffelt}. In the standard model 
such processes appear at the one-loop level. 
A typical example would be the diagram
shown in fig. \ref{ggnnbar} which contributes to the process
$\gamma\gamma\to \nu\bar\nu$.

The $2\to 2$ processes
$\gamma\gamma\to\nu\bar\nu$, $\gamma\nu\to\gamma\nu$
and $\nu\bar\nu\to\gamma\gamma$ were considered
already before the advent of the standard model
using the four-Fermi interaction \cite{chimor}.
However in the Fermi limit they vanish
due to the Landau-Yang theorem, as
was noted by Gell-Mann \cite{gellmann}
(for massless neutrinos, and with both photons on-shell).
In the standard model this suppression manifests itself
by factors of $\omega\over M_W$, where $\omega$ is the
center-of-mass energy and $M_W$ the $W$ boson mass.
Nonetheless the $2\to 2$ processes could be of importance
in very high energy reactions \cite{waxman,addr}, and
the one-loop helicity amplitudes for $\gamma\gamma\to\nu\bar\nu$
were calculated in \cite{addr}. 

There is no such suppression for processes involving two
neutrinos and more than two photons, which should therefore
be more important at low energies than the four-leg processes.
Following early work 
by Van Hieu and Shabalin \cite{hiesha}
recently the two-neutrino three-photon processes such as  
$\gamma\gamma \to \gamma\nu\bar\nu$,
$\gamma\nu \to \gamma\gamma\nu$,
$\nu\bar\nu \to \gamma\gamma\gamma$
have been investigated closely
\cite{dicrep,dikare,abmapi}.
For center-of-mass energies $2\omega$ between
$1$ keV and $1$ MeV the cross sections for those
$2\to 3$
processes indeed turn out to be larger than the
corresponding $2\to 2$ cross sections \cite{dicrep}.

Many more photon-neutrino processes become possible
if one admits neutrino masses or
anomalous magnetic dipole moments
\cite{ioaraf}.

In astrophysical environments it is often not realistic
to consider these processes as occuring in vacuum. Plasma
effects must be taken into account, 
as well as the presence of magnetic fields, 
which around pulsars can have field strengths surpassing
the critical value 
$B_{\rm crit} = {m_e^2\over e} =
4.41\times 10^{13}$ Gauss.
Of particular interest are then processes
which do not occur in vacuum but become possible
in a medium or B-field. An important example is
the plasmon decay $\gamma\to\nu\bar\nu$
\cite{adruwo,zaidi}, which is 
believed to be the dominant source for
neutrino production in many types of stars
\cite{raffelt}.
Similarly the Cherenkov process $\nu\to\nu\gamma$
becomes possible though it turns out to be of much
lesser astrophysical relevance \cite{galnik,ioaraf}.
In processes of this type the magnetic field plays a
double role. Firstly, it provides an effective
photon-neutrino coupling via intermediate
charged particles \cite{galnik,skobelev,demida}. 
Secondly, by modifying the photon dispersion relations
\cite{adler71,tsai74II,tsaerb,ditreu}
it opens up phase space for neutrino-photon reactions
of the type $1\to 2 + 3$.

Similarly one would expect the magnetic field 
to remove the Fermi limit
suppression of the above $2\to 2$ processes.
This has recently been verified both for the
weak \cite{shaisultanov97} and for the
strong field case \cite{chkumi}.

In the standard model
the effective coupling is provided by the diagrams
shown in fig. 2(a) and 2(b)
\footnote{I thank A.N. Ioannisian for providing this
figure.}. The double line represents the
electron propagator in the presence of the B-field.

\begin{figure}
\centering\leavevmode
\vbox{
\unitlength=0.8mm
\begin{picture}(60,25)
\put(8,15){\line(-1,1){8}}
\put(8,15){\line(-1,-1){8}}
\put(0,7){\vector(1,1){4}}
\put(8,15){\vector(-1,1){6}}
\multiput(9.5,15)(6,0){3}{\oval(3,3)[t]}
\multiput(12.5,15)(6,0){3}{\oval(3,3)[b]}
\put(31,15){\circle{10}}
\put(31,15){\circle{9}}
\multiput(37.5,15)(6,0){3}{\oval(3,3)[t]}
\multiput(40.5,15)(6,0){3}{\oval(3,3)[b]}
\put(0,10){\shortstack{{}$\nu$}}
\put(18,18){\shortstack{{$Z$}}}
\put(43,18){\shortstack{{$\gamma$}}}
\put(30,11){\shortstack{{e}}}
\put(57,13){\shortstack{{(a)}}}
\end{picture}

\unitlength=0.8mm
\begin{picture}(60,32)
\put(16,15){\line(-1,1){7.5}}
\put(16,15){\line(-1,-1){7.5}}
\put(15,15){\line(-1,1){7}}
\put(15,15){\line(-1,-1){7}}
\put(16,15){\line(-1,1){16}}
\put(16,15){\line(-1,-1){16}}
\put(1,0){\vector(1,1){6}}
\put(16,15){\vector(-1,1){10}}
\multiput(17.5,15)(6,0){3}{\oval(3,3)[t]}
\multiput(20.5,15)(6,0){3}{\oval(3,3)[b]}
\multiput(8.2,8.7)(0,6){3}{\oval(3,3)[l]}
\multiput(8.2,11.7)(0,6){2}{\oval(3,3)[r]}
\put(2,15){\shortstack{{}$W$}}
\put(0,3){\shortstack{{}$\nu$}}
\put(23,18){\shortstack{{$\gamma$}}}
\put(57,13){\shortstack{{(b)}}}
\end{picture}

\vskip3ex

\unitlength=0.8mm
\begin{picture}(60,25)
\put(8,15){\line(-1,1){8}}
\put(8,15){\line(-1,-1){8}}
\put(0,7){\vector(1,1){4}}
\put(8,15){\vector(-1,1){6}}
\put(13,15){\circle{10}}
\put(13,15){\circle{9}}
\multiput(19.5,15)(6,0){3}{\oval(3,3)[t]}
\multiput(22.5,15)(6,0){3}{\oval(3,3)[b]}
\put(0,10){\shortstack{{}$\nu$}}
\put(26,18){\shortstack{{$\gamma$}}}
\put(12,11){\shortstack{{e}}}
\put(57,13){\shortstack{{(c)}}}
\end{picture}
}
\smallskip
\caption[...]{Neutrino-photon coupling in an external magnetic field.
(a)~$Z$-$A$-mixing. (b)~Penguin diagram (only for $\nu_e$).
(c)~Effective coupling in the Fermi limit.
\label{Fig2}}
\end{figure}
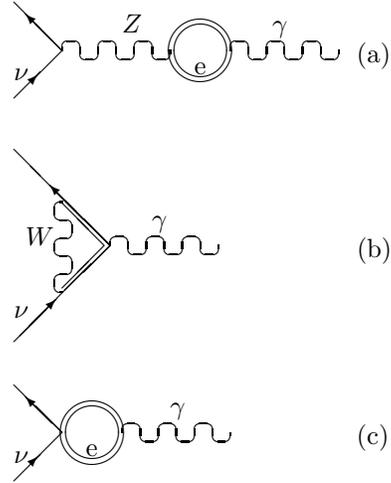
\no
In the limit of infinite gauge-boson masses
both diagrams can be replaced by diagram fig. 2(c).
The amplitude then effectively reduces
to a purely photonic amplitude with one of the
photons replaced by the neutrino current.
One is therefore led to the study of mixed
vector-axialvector amplitudes, or alternatively of
the corresponding effective action
\cite{dicrep,shaisultanov97,ioannisian}.

\section{Second-Order Fermions}

\no
Let us thus
consider the (Euclidean) one-loop effective action
for a Dirac fermion coupled to (abelian) vector and
axialvector background fields,

\be
\Gamma [A,A_5] = \ln {\rm Det}
[\slash p +e\slash A
+e_5\gamma_5{\slash A}_5
-im 
]
\label{defEAeuc}
\ee\no
It is easily shown 
that

\be
(\slash p + e\slash A +e_5\gamma_5 \not\!\!A_5)^2 =
- (\partial_\mu 
+ i{\cal A}_\mu)^2 + V
\label{introAcal}
\ee\no
where

\bear
{\cal A}_{\mu} &\equiv& e A_{\mu} -e_5\gamma_5\sigma_{\mu\nu}A_5^{\nu}
\non\\
V &\equiv& -i{e\over 2}
\sigma_{\mu\nu}
\Bigl(
\partial_{\mu}A_{\nu}-\partial_{\nu}A_{\mu}
\Bigr)
+ie_5\gamma_5A_{5,\mu}^{\mu} 
\non\\
&& + (D-2) e_5^2 A_5^2 
\label{defcalA,V}
\ear\no
($\sigma_{\mu\nu}=\half[\gamma_{\mu},\gamma_{\nu}]$).
We have used the four - dimensional Dirac algebra
relations,
but continued to $D$ dimensions
using an anticommuting $\gamma_5$.
Appealing to the usual argument that

\bear
&&{\rm Det}
\Bigl[(\not\!\!p +e \not\!\!A +\gamma_5 e_5\not\!\!A_5) - im\Bigr] 
\non\\ &=& 
{\rm Det}
\Bigl[(\not\!\!p + e \not\!\!A +\gamma_5 e_5 \not\!\!A_5) + im\Bigr] 
\non\\
&=& 
{\rm Det}^{1/2}
\Bigl[(\not\!\!p + e
\not\!\!A +\gamma_5 e_5 \not\!\!A_5)^2 + m^2\Bigr]
\non\\
\label{gamma5trick}
\ear\no
we can write
the effective action also in the following form,

\bear
\Gamma &=& -\half\, \Tr\, \int_0^\infty \, 
\frac{dT}{T} \, \exp 
\Bigl\lbrace
- T [-
(\partial_\mu + i{\cal A}_\mu)^2 
\non\\&&\hspace{90pt}
+ V + m^2 ]
\Bigr\rbrace\non\\
\label{Weucl}
\ear
Up to the global sign, this is 
formally identical with the effective action for
a scalar loop in a background containing a 
(Clifford algebra valued)
gauge field ${\cal A}$ and a potential $V$.
Note that the exponent is not hermitian. This 
distinguishes the present approach from previous ones
\cite{mnss,dhogag}, and makes it possible to
avoid the splitting of the effective action into its
real and imaginary parts.

For the pure vector (i.e. QED) case this representation of the
effective action is quite well-known. In this case
eq.(\ref{introAcal}) expresses the square of the Dirac operator
in terms of the Klein-Gordon operator with an additional
spin term $\sim \sigma_{\mu\nu}F^{\mu\nu}$. This has been used
to construct a non-standard ``second-order'' version of
spinor QED which is equivalent to the ordinary formulation,
but has a quite different set of Feynman rules 
\cite{feygel,hostler,morgan}. Those rules are shown in the
upper part of fig. 3
(in Euclidean conventions).

%%%%%%%%%%%%% cut here %%%%%%%%%%%%%%%%
\newlength{\fdwidth}
\setlength{\fdwidth}{.9in}
\newlength{\frwidth}
\setlength{\frwidth}{.9in}
\begin{figure}[t]
%\arraystretch{1.5}
\centering
% \begin{tabular}{cc@{\hspace{0.7in}}cc}
%\begin{tabular}{r@{\hspace{0.2in}}l@{\hspace{0.6in}}r@{\hspace{0.2in}}l}
\begin{tabular}{r@{\hspace{0.2in}}l}
\raisebox{0.2in}{\epsfig{file=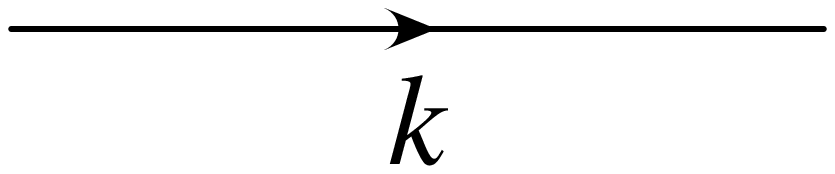, width=\fdwidth} } 
& \raisebox{0.65in}{\parbox{\frwidth}{\begin{displaymath} \frac{1}{k^2
+m^2} 
\end{displaymath}}} 
\vspace{-.21in}
\\
\epsfig{file=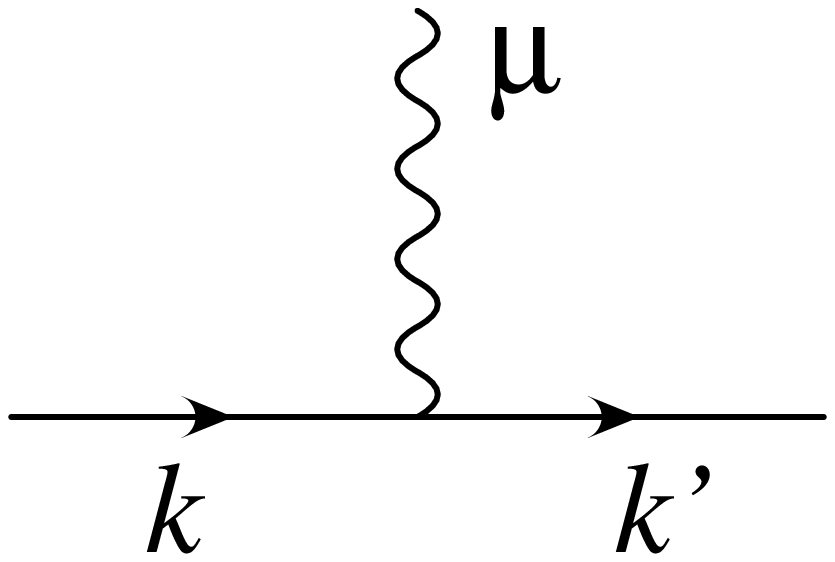, width=\fdwidth}   
& \raisebox{0.6in}{\parbox{\frwidth}{$$ e \left( k + k'\right)_\mu $$}} 
\\
\vspace{-.21in}
\epsfig{file=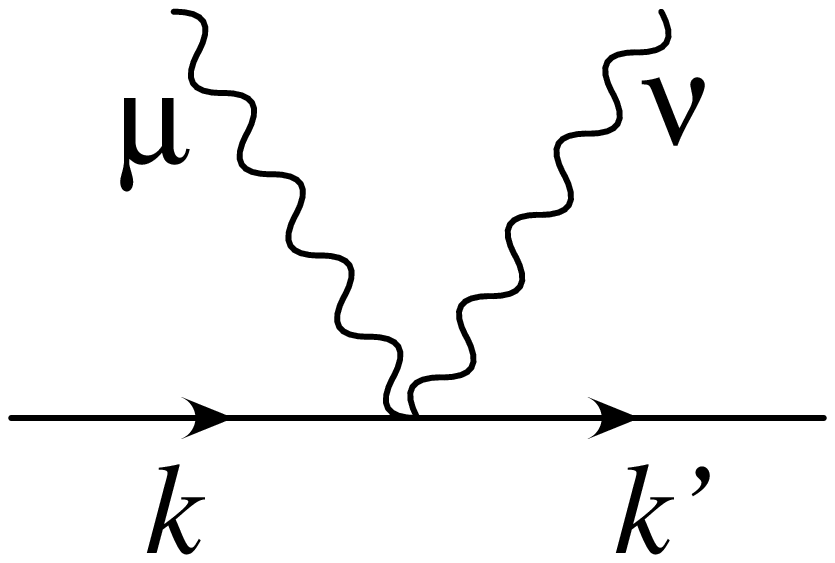,width=\fdwidth} 
& \raisebox{0.6in}{ \parbox{\frwidth}{ $$-2e^2g_{\mu\nu}$$ } }
\\\vspace{-.21in}
\epsfig{file=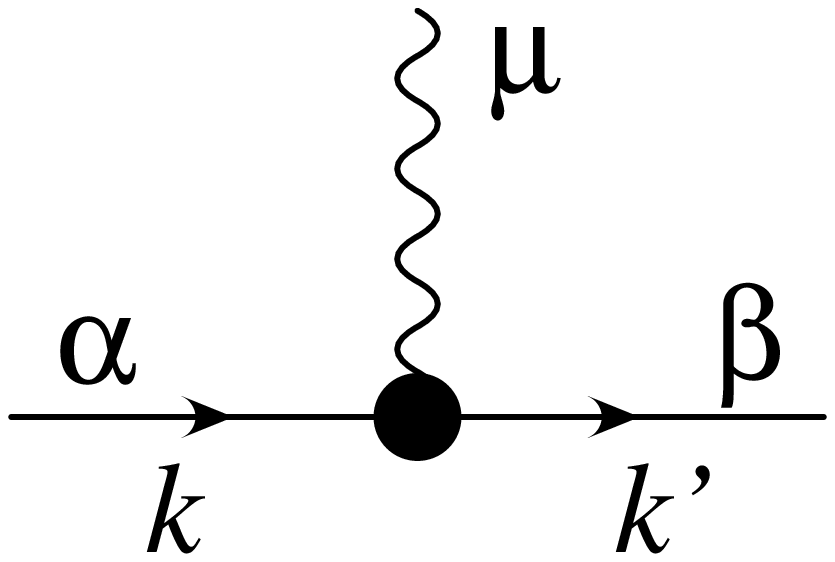,width=\fdwidth} 
& \raisebox{0.6in}{\parbox{\frwidth}{$$e (\sigma_{\mu\nu})_{\beta\alpha}\left
(k'-k\right )^\nu$$} }
\\\vspace{-.21in}
\epsfig{file=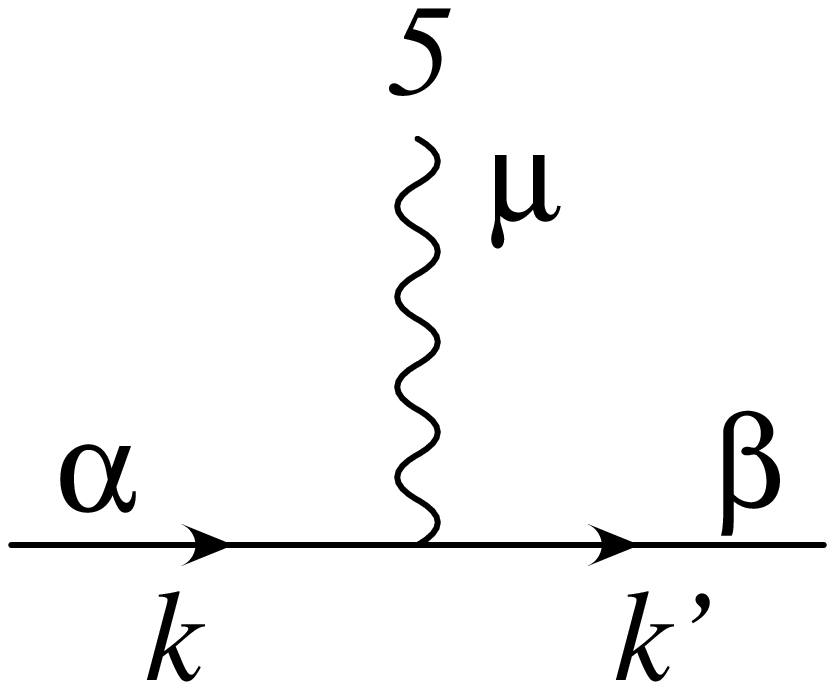, width=\fdwidth} 
& \raisebox{0.65in}{\parbox{\frwidth}{$$e_5 
(\gamma_5 \sigma_{\mu\nu})_{\beta\alpha}
\left( k + k' \right) ^ \nu $$ }}\\
 \epsfig{file=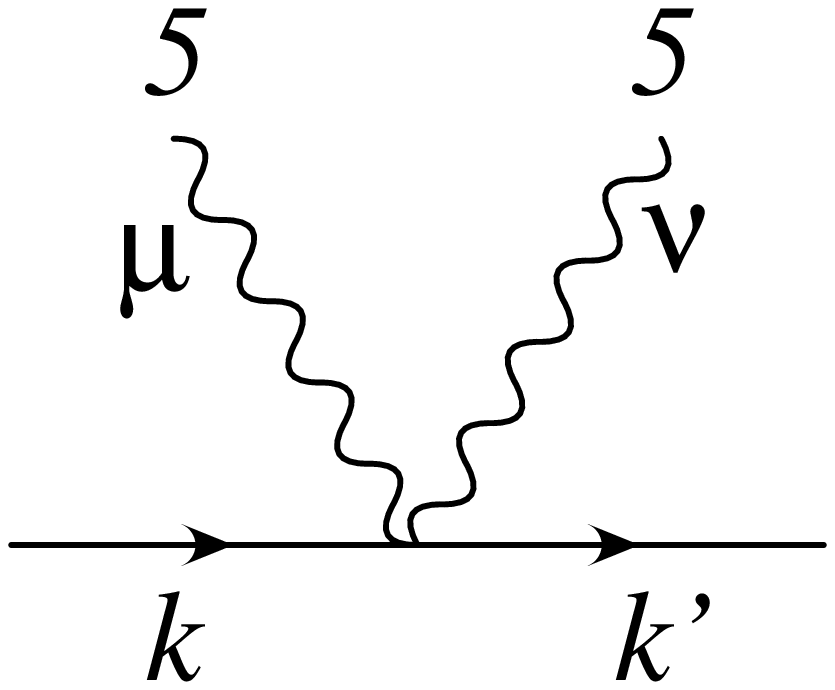,width=\fdwidth} 
& \raisebox{0.6in}{\parbox{\frwidth}{$$ 2 e_5^2 g_{\mu\nu}$$}}\\
\vspace{-.21in}
 \epsfig{file=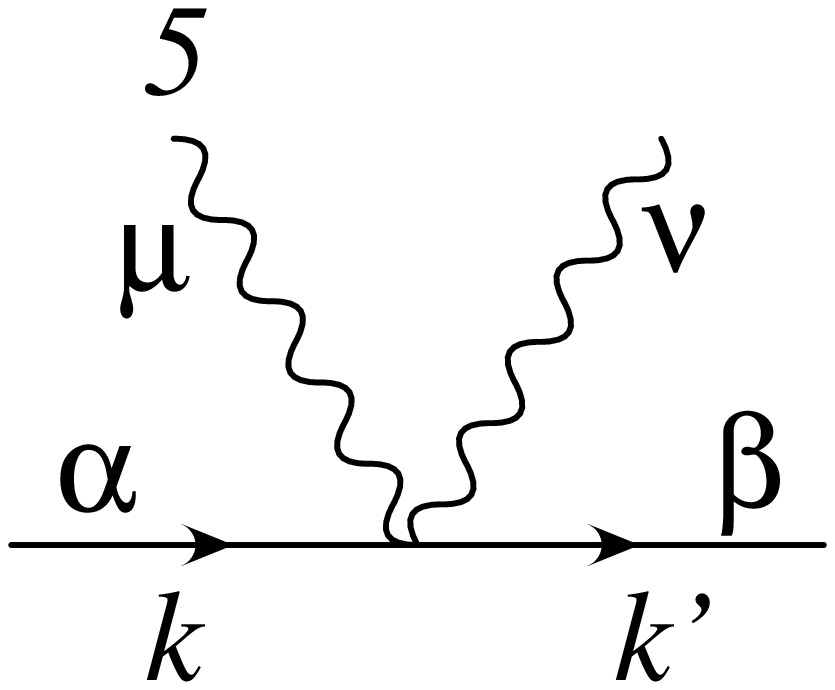,width=\fdwidth}
& \raisebox{0.6in}{\parbox{\frwidth}{$$-2 e e_5 (\gamma_5
\sigma_{\mu\nu}
)_{\beta\alpha}$$}}\\\vspace{-.21in}
\epsfig{file=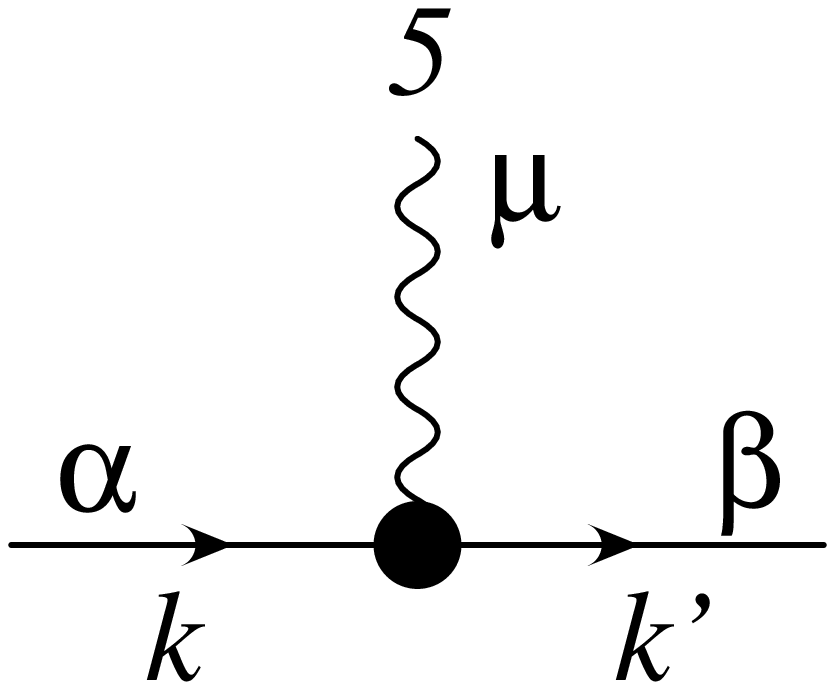,width=\fdwidth} 
& \raisebox{0.6in}{\parbox{\frwidth}{$$ e_5 
(\gamma_5)_{\beta\alpha}\left( k' - k \right)_\mu$$}}
\end{tabular}
\caption{Second order Feynman rules.}
\end{figure}
%%%%%%%%%%%%% cut here %%%%%%%%%%%%%%%%
\no
The electron propagator is of scalar form as shown in the first
line (the photon propagator is as usual). The first two vertices are
the same as in scalar QED, while the third one represents the
above spin term. Closed fermion loops require a spinor trace and a
factor of $-\half$. For the treatment of external fermions in the
second order
formalism see \cite{morgan}. 

In the presence of both vectors and axialvectors there are four
more vertices shown in the lower part of fig. 3. Those can also be 
directly read off eq.(\ref{introAcal}). 
Again these rules are guaranteed to yield the same amplitudes
as in the usual first-order formalism, but rewritten in
a very different Feynman diagram expansion.

\section{Path Integral Representation of
Vector -- Axialvector Couplings}

Rather than applying the second-order formalism directly, 
we use it for the derivation of a worldline
path integral suitable for vector -- axialvector calculations
\cite{mcksch,dimcsc}. Using standard coherent state methods
\cite{ohnkas} the functional trace in our representation
eq.(\ref{Weucl}) for the effective action can be transformed
into the following quantum mechanical double path integral,

\bear
\Gamma &=&
-\half
\int_0^{\infty}{dT\over T}
\,\e^{-m^2T}
\Dx
\int
{\cal D}\psi
\,\,\e^{-\int_0^Td\tau\, L(\tau)}\non\\
L &=& 
\kinb + \half\psi\cdot\dot\psi 
+ie\dot x\cdot A
-ie \psi\cdot F\cdot\psi
\nonumber\\&&
 +ie_5\hat\gamma_5
\Bigl[
-2\dot x\cdot\psi\psi\cdot A_5
+\partial\cdot A_5
\Bigr]
\non\\&&
+ (D-2) e_5^2A_5^2
\label{defL}
\ear\no
Here $T$ denotes the usual Schwinger proper-time
parameter for the fermion circulating in the loop.
The ``coordinate'' path integral
$\int{\cal D} x$ has to be performed over the space
of closed loops in spacetime $x^{\mu}(\tau )$ with
period $T$, $x^{\mu}(T) = x^{\mu}(0)$.
The second path integral $\int{\cal D}\psi$,
which takes the fer\-mion spin into account,  
is to be integrated over the space of
Grassmann-valued functions
$\psi^{\mu}(\tau )$. 
The boundary conditions on the Grassmann path integral are,
after expansion of the interaction exponential, determined by the
power of $\hat\gamma_5$ appearing in a given term; they are
(anti) periodic with period $T$ if that power is even (odd).
After the boundary conditions are determined $\hat\gamma_5$ can be
replaced by unity.
Comparing (\ref{defL}) with (\ref{Weucl})
one notes that the coherent state formalism has
produced a  formal correspondence
$[\gamma^{\mu},\gamma^{\nu}] \rightarrow 4 
\psi^{\mu}\psi^{\nu}$, 
$\gamma_5 \rightarrow \hat \gamma_5$.

We proceed to the perturbative calculation of this
path integral following the recipes of the
``string-inspired formalism''.
Before tackling the full vector -- axialvector case
it will, however, be useful to first study
the simpler cases of the Scalar QED and Spinor QED
photon scattering amplitudes.

\subsection{Scalar Quantum Electrodynamics}

For the case of scalar QED, the analogue of (\ref{defL})
is well-known and goes, in fact, back to Feynman \cite{feyn}.
The one-loop effective action due to a scalar loop for a Maxwell
background can be written as

\bear
\Gamma[A]
&=&
\Tintm
\int_{x(T)=x(0)}{\cal D}x(\tau)
\non\\&&\times
\e^{-\int_0^T d\tau\Bigl(
\kinb
+ie\,\dot x\cdot A(x(\tau))
\Bigr)}
\label{scalarqedpi}
\ear\no
If we expand the
``interaction exponential'',

\bear
{\rm exp}\Bigl[
-\int_0^Td\tau\, ieA_{\mu}\dot x^{\mu}
\Bigr]
&=&\sum_{N=0}^{\infty}
{{(-ie)}^N\over N!}
\prod_{i=0}^N
\int_0^Td\tau_i
\non\\&&\hspace{-20pt}
\times
\biggl[
\dot x^{\mu}(\tau_i)
A_{\mu}(x(\tau_i))
\biggr]
\label{expandint}
\ear\no
the individual terms correspond to Feynman diagrams
describing a fixed number of
interactions of the scalar loop with
the external field (fig. 4).

%\vspace{-5pt}
\begin{figure}[t]
\begin{center}
\begin{picture}(-100,0)%
\epsfxsize10cm
\epsfxsize5cm
\epsfig{file=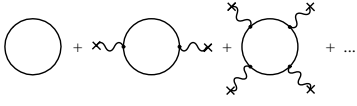}%

\end{picture}%
\setlength{\unitlength}{0.00087500in}%
\begingroup\makeatletter\ifx\SetFigFont\undefined
% extract first six characters in \fmtname
\def\x#1#2#3#4#5#6#7\relax{\def\x{#1#2#3#4#5#6}}%
\expandafter\x\fmtname xxxxxx\relax \def\y{splain}%
\ifx\x\y   % LaTeX or SliTeX?
\gdef\SetFigFont#1#2#3{%
  \ifnum #1<17\tiny\else \ifnum #1<20\small\else
  \ifnum #1<24\normalsize\else \ifnum #1<29\large\else
  \ifnum #1<34\Large\else \ifnum #1<41\LARGE\else
     \huge\fi\fi\fi\fi\fi\fi
  \csname #3\endcsname}%
\else
\gdef\SetFigFont#1#2#3{\begingroup
  \count@#1\relax \ifnum 25<\count@\count@25\fi
  \def\x{\endgroup\@setsize\SetFigFont{#2pt}}%
  \expandafter\x
    \csname \romannumeral\the\count@ pt\expandafter\endcsname
    \csname @\romannumeral\the\count@ pt\endcsname
  \csname #3\endcsname}%
\fi
\fi\endgroup
\begin{picture}(3634,1536)(20,-850)
\end{picture}
\caption{\label{fig3} 
Expanding the path integral in powers of the
background field.}
\end{center}
\end{figure}
%\vspace{-5pt}

\no
The corresponding $N$ -- photon
scattering amplitude is then obtained by
specializing to a background
consisting of 
a sum of plane waves with definite
polarizations,

\be
A_{\mu}(x)=
\sum_{i=1}^N
\varepsilon_{i\mu}
\e^{ik_i\cdot x}
\label{planewavebackground}
\ee\no
and picking out the term containing every
$\varepsilon_i$ once.
This yields the following representation for
the $N$ - photon amplitude, 

\bear
\Gamma[\lbrace k_i,\varepsilon_i\rbrace\rbrace]
&=&
(-ie)^{N}
\Tintm 
\PITD
\non\\&&
\hspace{-35pt}
\times
\Bigl\langle
V_{A}^{0}[k_1,\varepsilon_1]\ldots
V_{A}^{0}[k_N,\varepsilon_N]
\Bigr\rangle
\non\\
\label{repNvector}
\ear\no
Here $V_A^{0}$ denotes 
the same photon
vertex operator as is used in string perturbation
theory \cite{grscwi},

\begin{equation}
V_A^{0}[k,\varepsilon]
\equiv
\int_0^Td\tau\,
\varepsilon\cdot \dot x(\tau)
\,{\rm e}^{ikx(\tau)}
\label{photonvertopscal}
\end{equation}
\noindent
At this stage the path integral has become Gaussian,
which reduces its evaluation to the task
of Wick contracting the expression

\be
\biggl\langle
\dot x_1^{\mu_1}\e^{ik_1\cdot x_1}
\cdots
\dot x_N^{\mu_N}\e^{ik_N\cdot x_N}
\biggr\rangle
\label{scalqedwick}
\ee
\no
The Green's function to be used is 
simply the one for the second-derivative
operator, acting on
periodic functions. To derive 
this Green's function, first observe that 
$\int{\cal D}x(\tau)$ contains the constant functions,
which must be removed to obtain a well-defined
inverse. One therefore restricts the integral
over the space of all loops by fixing the average
position $x_0^{\mu}$ of the loop,

\begin{equation}
x_0^{\mu}\equiv {1\over T}\int_0^T d\tau\, x^{\mu}(\tau)
\label{defx0}
\end{equation}
\no
For effective action calculations this reduces the
effective action to the effective Lagrangian.
In scattering amplitude calculations,
the integral over $x_0$ just gives
momentum conservation.
The reduced path integral $\int{\cal D}y(\tau)$
over $y(\tau)\equiv x(\tau) - x_0$
has an invertible kinetic operator. This
inverse is 

\be
2\bigl\langle\tau_1\mid
{\Bigl({d\over d\tau}\Bigr)}^{-2}
\mid\tau_2\bigr\rangle
=
G_B(\tau_1,\tau_2)
\label{calcG}
\ee

\no
with the ``bosonic'' worldline Green's function

\be
G_B(\tau_1,\tau_2)=\mid \tau_1-\tau_2\mid 
-{{(\tau_1-\tau_2)}^2\over T} 
\label{defGB}
\ee
\no
For the performance of the Wick contractions it is
convenient to formally exponentiate all the $\dot x_i$'s, 
writing

\be
\varepsilon_i\cdot
\dot x_i\e^{ik_i\cdot x_i}
=
\e^{\varepsilon_i\cdot\dot x_i
+ik_i\cdot x_i}
\mid_{{\rm lin}(\varepsilon_i)}
\label{formexp}
\ee
\no
This allows one to rewrite the product of $N$ photon vertex
operators as an exponential. Then one needs only
to ``complete the square'' to arrive at the following
closed expression for the one-loop
$N$ - photon amplitude,

\begin{eqnarray}
\Gamma[k_1,\varepsilon_1;\ldots;k_N,\varepsilon_N]
&=&
{(-ie)}^N
{(2\pi )}^D\delta (\sum k_i)
\non\\
&&\hspace{-80pt}
\times {\dps\int_{0}^{\infty}}{dT\over T}
{[4\pi T]}^{-{D\over 2}}
e^{-m^2T}
\prod_{i=1}^N \int_0^T 
d\tau_i
\nonumber\\
&&\hspace{-80pt}
\times
\exp\biggl\lbrace\sum_{i,j=1}^N 
\bigl\lbrack \half G_{Bij} k_i\cdot k_j
+i\dot G_{Bij}k_i\cdot\varepsilon_j 
\nonumber\\
&&\hspace{-40pt}
+\half\ddot G_{Bij}\varepsilon_i\cdot\varepsilon_j
\bigr\rbrack\biggr\rbrace
\mid_{\rm multi-linear}.
\nonumber\\
\label{scalarqedmaster}
\end{eqnarray}
\no
Here it is understood that only the terms linear
in all the $\varepsilon_1,\ldots,\varepsilon_N$
have to be taken. 
Besides the Green's function $G_B$ also its first and
second deriatives appear,

\begin{eqnarray}
\dot G_B(\tau_1,\tau_2) &=& {\rm sign}(\tau_1 - \tau_2)
- 2 {{(\tau_1 - \tau_2)}\over T}\nonumber\\
\ddot G_B(\tau_1,\tau_2)
&=& 2 {\delta}(\tau_1 - \tau_2)
- {2\over T}\quad \nonumber\\
\label{GdGdd}
\end{eqnarray}
\noindent
Dots generally denote a
derivative acting on the first variable,
$\dot G_B(\tau_1,\tau_2) \equiv {\partial\over
{\partial\tau_1}}G_B(\tau_1,\tau_2)$, 
and we abbreviate
$G_{Bij}\equiv G_B(\tau_i,\tau_j)$ etc.
The factor ${[4\pi T]}^{-{D\over 2}}$
represents the free Gaussian path integral
determinant.

The expression (\ref{scalarqedmaster})
is identical with the ``Bern-Kosower
Master Formula'' 
for the special case considered
\cite{berkos}.
Let us consider explicitly the vacuum polarisation case,
$N=2$. For $N=2$ the expansion of the exponential
factor 
yields the following expression,

\be
\Bigl(
\ddot G_{B12}\varepsilon_1\cdot\varepsilon_2
+
\dot G_{B12}^2
\varepsilon_1\cdot k_2
\varepsilon_2\cdot k_1
\Bigr)
{\rm e}^{G_{B12}k_1\cdot k_2}
\non\\
\label{N=2wick}
\ee\no
Now one could just plug in the explicit
formulas from eqs.(\ref{defGB}),(\ref{GdGdd}) and
do the integrals. It is useful, however,
to first perform
a partial integration on the first
term of eq. (\ref{N=2wick}) in either
$\tau_1$ or $\tau_2$. The integrand then
turns into

\be
\Bigl(
\varepsilon_1\cdot\varepsilon_2
k^2
-
\varepsilon_1\cdot k
\varepsilon_2\cdot k
\Bigr)
\dot G_{B12}^2
{\rm e}^{-G_{B12}k^2}
\label{N=2partint2}
\ee\no
($k=k_1=-k_2$).
Thus the partial integration leads to the
appearance of a transversal projector, making
the transversality of the
vacuum polarization amplitude manifest at the
parameter integral level.
We rescale to the unit circle, 
$\tau_i = Tu_i, i = 1,2$, and use translation
invariance in $\tau$ to fix the zero to 
be at the location of the second vertex operator,
$u_2=0, u_1=u$.
We have then

\bear
G_B(\tau_1,\tau_2)&=&Tu(1-u)\nonumber\\
\dot G_B(\tau_1,\tau_2)&=&1-2u
\nonumber\\
\label{scaledown}
\ear\no
After performing the trivial $T$ - integration
one arrives at

\bear
\Gamma^{\mu\nu}_{\rm scal}
[k] &=&
{e^2\over {(4\pi )}^{D\over 2}}
\Bigl(k^{\mu}k^{\nu}-g^{\mu\nu}k^2\Bigr)
\Gamma\bigl(2-{D\over 2}\bigl)
\non\\&&\hspace{-26pt}
\times
\int_0^1du
(1-2u)^2
{\Bigl[
m^2 + u(1-u)k^2
\Bigr]
}^{{D\over 2}-2}
\non\\
\label{scalarvpresult}
\ear\no
The result of the final integration is, of course,
the same as is found
in the standard field theory calculation
of the corresponding two Feynman diagrams.

\subsection{Spinor Quantum Electrodynamics}

For spinor QED, the worldline representation of the
one-loop effective action is just eq.(\ref{defL})
with $A_5 = 0$,

\begin{eqnarray}
\Gamma\lbrack A\rbrack &  = &- \half {\displaystyle\int_0^{\infty}}
{dT\over T}
e^{-m^2T}
{\displaystyle\int} 
{\cal D} x
{\displaystyle\int}
{\cal D}\psi\nonumber\\
& \phantom{=}
&\times
{\rm exp}\biggl [- \int_0^T d\tau
\Bigl ({1\over 4}{\dot x}^2 + {1\over
2}\psi\dot\psi
\non\\&&
+ ieA\cdot x - ie
\psi\cdot F \cdot\psi
\Bigr )\biggr ]
\label{spinorpi}
\end{eqnarray}

\noindent
The
calculation of the $x$ - path integral
proceeds as before.
Concerning the $\psi$ - path integral,
first note that $\psi$ is antiperiodic 
in the absence of $\hat\gamma_5$,
so that there is no zero mode.
To find the appropriate ``fer\-mio\-nic''
worldline Green's function $G_F$,
we thus need to invert the first 
derivative in the space of
anti-periodic functions. This
yields 

\be
2\bigl\langle\tau_1\mid
{({d\over d\tau})}^{-1}
\mid\tau_2\bigr\rangle
=
{\rm sign}(\tau_1-\tau_2)
\equiv G_F(\tau_1,\tau_2)
\label{calcGF}
\ee
\no
Thus we have now the following two Wick contraction
rules,

\begin{eqnarray}
\langle y^{\mu}(\tau_1)y^{\nu}(\tau_2)\rangle
   & = &- g^{\mu\nu}G_B(\tau_1,\tau_2)
\nonumber\\
\langle \psi^{\mu}(\tau_1)\psi^{\nu}(\tau_2)\rangle
   & = &{1\over 2}\, g^{\mu\nu} G_F(\tau_1,\tau_2)
\nonumber\\
\label{wickrules}
\end{eqnarray}
\noindent
With our conventions the free 
$\psi$ - path integral
is normalized as

\begin{eqnarray}
N_A \equiv
{\displaystyle\int} {\cal D} \psi\,
{\rm exp}\Bigl [- \int_0^T d\tau
{1\over2}\psi\dot\psi\Bigr ]
& = & 4\qquad \nonumber\\
\label{norm}
\end{eqnarray}
\noindent
The photon vertex operator $V_A$
now has an additional Grassmann
piece
representing the coupling of the
photon to the spin degree of freedom
in the loop,

\begin{equation}
V_A^{\half}[k,\varepsilon]
\equiv
\int_0^Td\tau
\Bigl(
\varepsilon\cdot \dot x
+2i
\varepsilon\cdot\psi
k\cdot\psi
\Bigr)\,
{\rm e}^{ikx}
\label{photonvertopspin}
\end{equation}
\noindent
Let us again look at the vacuum polarization case, $N=2$.
We need now to Wick-contract two copies of the
above vertex operator.
The calculation
of $\int{\cal D}x$ is identical with the scalar QED calculation.
Only the calculation of $\int{\cal D}\psi$ is new, and amounts to
a single Wick contraction,

\be
{(2i)}^2
\Bigl\langle
\psi^{\mu}_1\psi_1\cdot k_1
\psi^{\nu}_2\psi_2\cdot k_2
\Bigr\rangle
=-
G_{F12}^2
\Bigl[g^{\mu\nu}k^2-k^{\mu}k^{\nu}\Bigr]
\label{spinwick2point}
\ee\no
Note that,
up to the global normalization, 
the parameter integral for the spinor loop is obtained
from the one for the scalar loop simply by substituting, in
eq.(~\ref{N=2partint2}),

\be
\dot G_{B12}^2 \rightarrow
\dot G_{B12}^2 
- G_{F12}^2 =-{4\over T}G_{B12}
\label{subs2point}
\ee\no
The complete change thus amounts
to supplying eq.(~\ref{scalarvpresult})
with the global factor of $-2$, and replacing ${(1-2u)}^2$
by $-4u(1-u)$. This leads to

\bear
\Gamma^{\mu\nu}_{\rm spin}[k]
&=&
8{e^2\over {(4\pi )}^{D\over 2}}
\Bigl(k^{\mu}k^{\nu}-g^{\mu\nu}k^2\Bigr)
\Gamma\bigl(2-{D\over 2}\bigl)
\non\\
&&\hspace{-26pt}
\times
\int_0^1du\,
u(1-u)
{\Bigl[
m^2 + u(1-u)k^2
\Bigr]
}^{{D\over 2}-2}
\non\\
\label{spinorvpresult} 
\ear\no
again in agreement with field theory.
The above ``substitution rule''
carries over to arbitrary $N$ as follows
\cite{berkos}.
Up to the global normalization factor,
the integrand for the spin - $\half$ case
can be obtained from the spin - $0$
integrand simply by replacing,
after the partial integration
procedure, every
``cycle'' 
$\dot G_{Bi_1i_2} 
\dot G_{Bi_2i_3} 
\cdots
\dot G_{Bi_ni_1}$
by 

\begin{eqnarray}
\dot G_{Bi_1i_2} 
\dot G_{Bi_2i_3} 
\cdots
\dot G_{Bi_ni_1}
&\rightarrow &
{\rm same}\hfill
\nonumber\\
&&\hspace{-60pt}
-
G_{Fi_1i_2}
G_{Fi_2i_3}
\cdots
G_{Fi_ni_1}\nonumber\\
\label{subrule}
\end{eqnarray}

\noindent
This rule
reduces the transition from the
scalar loop case to the spinor loop
case to a pattern matching problem.

\subsection{The General Case}

We are now ready for the general vector -- axialvector
case.
The only real novelty is in the case
of an odd number of axial vectors, since here
the Grassmann path integral 
appears with periodic boundary conditions.
It acquires therefore a zero  mode which
must be separated out in the same way 
as for the coordinate path
integral, splitting

\bear
\psi^{\mu}(\tau) &=& \psi_0^{\mu} + \xi^{\mu}(\tau)
\label{splitgrass}\\
\int_0^Td\tau \, \xi^{\mu}(\tau) &=& 0
\label{xicond}
\ear\no
The zero mode integration then produces the
$\varepsilon$ - tensor expected for a spinor loop
with an odd number of axial insertions,

\be
\int d^4\psi_0
\psi_0^{\mu_1}\psi^{\mu_2}_0\psi^{\mu_3}_0\psi^{\mu_4}_0
=\varepsilon^{\mu_1\cdots\mu_4}
\label{zeromodeintegral}
\ee\no
The remaining $\xi$ - path integral 
is again performed using the appropriate Wick contraction
rule, which is

\begin{eqnarray}
\langle\xi^{\mu}(\tau_1)\, \xi^{\nu}(\tau_2)
&=& 
g^{\mu\nu}\half {\dot{G}}_{B}(\tau_1,\tau_2)
\qquad
\label{xicorr}
\end{eqnarray}\no
The free Grassmann path integral with
periodic boundary conditions is normalized to 
$N_P = 1$.

To define the analogue of the photon vertex operator 
(\ref{photonvertopspin})
for the axial coupling, it is convenient to linearize the
term quadratic in $A_5$. 
This can be achieved through the introduction
of an auxiliary path
integration, writing

\bear
\label{linearizeA52}
\exp \Bigl[-(D-2)e_5^2\int_0^Td\tau A_5^2\Bigr]
&=&
\int {\cal D}z 
\non\\
&&
\hspace{-100pt}\times 
\exp \Bigl[-\int_0^Td\tau 
\Bigl({z^2\over 4}+ie_5\sqrt{D-2}z\cdot A_5\Bigr)
\Bigr]
\non\\
\ear\no
This allows us to define an axial-vector vertex operator
as follows,

\bear
V_{A_5}[k,\varepsilon] &\equiv&
\hat\gamma_5
\int_0^Td\tau
\Bigl(i\varepsilon\cdot k 
+ 2\varepsilon\cdot\psi
\dot x\cdot\psi
\non\\&&
\hspace{20pt}
+ \sqrt{D-2}\,\varepsilon\cdot z
\Bigr)
\,\e^{ik\cdot x}
\non\\
\label{defaxvectvertop}
\ear\no
With this definition
we obtain the following 
representation for the
one-loop amplitude with $M$ vectors and $N$ axialvectors,

\bear
\Gamma[\lbrace k_i,\varepsilon_i\rbrace,
\lbrace k_{5j},\varepsilon_{5j}\rbrace]
&=&
-\half N_{A,P}(-i)^{M+N}e^Me_5^N
\non\\&&
\hspace{-95pt}
\times
\Tintm 
\PITD
\Bigl\langle
V_{A}^{\half}[k_1,\varepsilon_1]\ldots
\non\\&&
\hspace{-95pt}
\ldots
V_{A}^{\half}[k_M,\varepsilon_M]
V_{A_5}[k_{51},\varepsilon_{51}]\ldots
V_{A_5}[k_{5N},\varepsilon_{5N}]
\Bigr\rangle
\non\\
\label{repMvectorNaxial}
\ear\no
The Wick-contraction of this expression can still be done
in closed form \cite{dimcsc},
though the result is too lengthy to be given here.

\section{Relation to the ``$\gamma_5$ - Problem''}

As is well-known, theories involving both vector
and axialvector couplings can, in the presence of
UV divergences, generally not be
regularised in a way which would respect both
the vector and the axialvector gauge invariance.
The most elementary manifestation of this
fact is the ABJ anomaly \cite{adler,beljac}.
Since the worldline formalism is normally used
together with dimensional regularization,
we should investigate how our formalism
behaves with respect to the chiral symmetry
in the dimensional continuation.

It must be remembered that,
in field theory, one has essentially a choice between
two evils. If one preserves the anticommutation
relations between $\gamma_5$ and the other Dirac matrices
\cite{chfuhi}
then the chiral symmetry is preserved for parity-even
fermion loops, but Dirac traces with an odd number
of $\gamma_5$'s are not unambiguously defined in general,
requiring additional prescriptions. The main alternative
is to use the
't Hooft-Veltman-Breitenlohner-Maison prescription
\cite{HV,Bremai}. In this case there are no ambiguities, but
the chiral symmetry is explicitly broken, so that 
in chiral gauge theories finite renormalizations
generally become
necessary to avoid violations of the gauge Ward identities
\cite{bonneau,trueman}.

Since our path integral representation was derived using an
anticommuting $\gamma_5$, we have not broken the chiral
symmetry. In particular, in the massless case the amplitude
with an even number of axialvectors 
should coincide with the
corresponding vector amplitude. 
It is easy to verify this fact for the two-point case.
The Wick contraction of two axialvector vertex operators
(\ref{defaxvectvertop}) leads, after rescaling
$\tau_{1,2}=T u_{1,2}$ and putting $u_2 =0$ as above,
 to the following expression for the
(massless) two-point amplitude, 

\bear
\Gamma^{\mu\nu}[k]
&=&
2\Tint \PITD 
\non\\&&\hspace{-34pt}
\times
\biggl\lbrace
2(D-2)Tg^{\mu\nu}
-2(D-1)Tg^{\mu\nu}
\non\\&&\hspace{-34pt}
+ \int_0^1 du
\,\e^{-Tu(1-u)k^2}
\Bigl[
2(D-1)Tg^{\mu\nu}
\non\\&&\hspace{-34pt}
+(1-2u)^2T^2(g^{\mu\nu}k^2-k^{\mu}k^{\nu})
+T^2k^{\mu}k^{\nu}
\Bigr]
\biggr\rbrace
\non\\
\label{A5A5int2}
\ear\no
The result of the integrations is,
using
$\Gamma$ - function identities,
easily identified with the
one for the
massless QED vacuum polarisation,
eq.(\ref{spinorvpresult}).
This confirms that the chiral symmetry is unbroken for
parity-even loops. 

\begin{figure}[ht]
\begin{center}
\begin{picture}(0,0)%
\epsfig{file=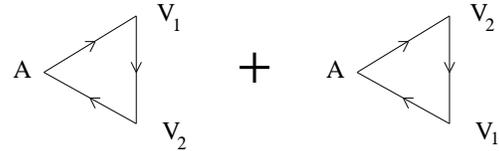}%
\end{picture}%
\setlength{\unitlength}{0.00047500in}%
\begingroup\makeatletter\ifx\SetFigFont\undefined
% extract first six characters in \fmtname
\def\x#1#2#3#4#5#6#7\relax{\def\x{#1#2#3#4#5#6}}%
\expandafter\x\fmtname xxxxxx\relax \def\y{splain}%
\ifx\x\y   % LaTeX or SliTeX?
\gdef\SetFigFont#1#2#3{%
  \ifnum #1<17\tiny\else \ifnum #1<20\small\else
  \ifnum #1<24\normalsize\else \ifnum #1<29\large\else
  \ifnum #1<34\Large\else \ifnum #1<41\LARGE\else
     \huge\fi\fi\fi\fi\fi\fi
  \csname #3\endcsname}%
\else
\gdef\SetFigFont#1#2#3{\begingroup
  \count@#1\relax \ifnum 25<\count@\count@25\fi
  \def\x{\endgroup\@setsize\SetFigFont{#2pt}}%
  \expandafter\x
    \csname \romannumeral\the\count@ pt\expandafter\endcsname
    \csname @\romannumeral\the\count@ pt\endcsname
  \csname #3\endcsname}%
\fi
\fi\endgroup
\begin{picture}(4834,2736)(180,-1050)
\end{picture}
\caption{\label{figtriangle}
Sum of AVV triangle diagrams.}
\end{center}
\end{figure}

Next let us see whether we correctly get the AVV anomaly,
i.e. the anomalous divergence for the
$\langle A_5AA\rangle$
amplitude. In field theory this amplitude would be given by
the two triangle diagrams shown in fig. \ref{figtriangle}.

\no
According to eq.(\ref{repMvectorNaxial}) the sum of these
diagrams is given by
\bear
\langle A^{\mu}[k_1]A^{\nu}[k_2]A_5^{\rho}[k_3]\rangle
&=&
-{i\over 2}
\Tintm
\non\\&&\hspace{-90pt}\times
\Dx\Dpsi
\,\,\exp
\biggl\lbrace
-\int_0^Td\tau
\,
\Bigl(
\fourth
\dot x^2 + \half \psi\cdot\dot\psi
\Bigr)
\biggr\rbrace
\non\\
&&\hspace{-90pt}\times
\int_0^Td\tau_1
\Bigl(\dot x^{\mu}_1+2i\psi^{\mu}_1
k_1\cdot\psi_1
\Bigr)
\e^{ik_1\cdot x_1}
\non\\
&&\hspace{-90pt}\times
\int_0^Td\tau_2
\Bigl(\dot x_2^{\nu}+2i\psi^{\nu}_2
k_2\cdot\psi_2
\Bigr)
\e^{ik_2\cdot x_2}
\non\\
&&\hspace{-90pt}\times
\int_0^Td\tau_3
\Bigl(
ik_3^{\rho}
+2\psi^{\rho}_3\dot x_3\cdot\psi_3
\Bigr)
\e^{ik_3\cdot x_3}
\label{AAA5amplitude}
\ear

From this expression it is obvious, even before doing any
integrations, that the amplitude will be
transversal in the vector current indices. If one multiplies
the right hand side by, say, $k_1^{\mu}$, then the Grassmann part
of the $\tau_1$ - integrand will vanish by the
anticommuation rules, and the remaining bosonic part of the photon vertex
operator becomes a total derivative in $\tau_1$, which vanishes
upon integration
due to periodicity. This mechanism is, of course, well-known from
string theory. Nothing analogous holds true for the axial-vector
vertex operator.

Thus the structure of the path integral eq.(\ref{defL})
already forces the  
vector currents to be divergence-free,
and we clearly have to look at the
axial vector current to find the anomalous divergence.
We are
interested in this divergence only, rather than in a calculation
of the complete amplitude,
so that we can simplify
by contracting eq.(\ref{AAA5amplitude}) with $k_3^{\rho}$.
Also we can put legs $1$ and $2$ on-shell,
$k_1^2=k_2^2=0$, and it suffices to consider the massless
case.

Since for this amplitude the Grassmann path integral has periodic
boundary conditions, according to eq.(\ref{splitgrass}) we have
to rewrite 

$$\psi_i^{\alpha}(\tau) =\psi_{0i}^{\alpha} + \xi_i^{\alpha}(\tau)$$

\no
and then to keep only those terms which contain 
precisely four factors of
the zero mode piece $\psi_0$. Using the zero mode integration
rule eq.(\ref{zeromodeintegral}) as
well as 
eq.(\ref{xicorr}) and the usual correlator for the
coordinate field, we obtain (deleting the
energy-momentum conservation factor)

\bear
k_3^{\rho}
\langle A^{\mu}A^{\nu}A_5^{\rho}\rangle
&=&
2
\varepsilon^{\mu\nu\kappa\lambda}k_1^{\kappa}k_2^{\lambda}
\Tint
{(4\pi T)}^{-2}
\non\\
&&\hspace{-60pt}\times 
\prod_{i=1}^3\int_0^Td\tau_i
\exp\Bigl\lbrack
(G_{B12}-G_{B13}-G_{B23})
k_1\cdot k_2
\Bigr\rbrack
\non\\
&&\hspace{-60pt}\times
\biggl\lbrace
\Bigl[2+
(\dot G_{B12} +\dot G_{B23} +\dot G_{B31})
(\dot G_{B13}-\dot G_{B23})
\Bigr]
\non\\
&&\hspace{-35pt}\times 
k_1\cdot k_2
-(\ddot G_{B13} +\ddot G_{B23})
\biggr\rbrace
\label{divAAA5}
\ear
Momentum conservation has been used to eliminate
$k_3$.
Removing the second derivatives
$\ddot G_{B13}$ ($\ddot G_{B23}$) by a
partial integration in $\tau_1$ ($\tau_2$), 
and using the identities

\bear
\dot G_{Bij}^2 &=& 1-4{G_{Bij}\over T}
\non\\
\dot G_{Bij} + \dot G_{Bjk} + \dot G_{Bki}
&=&
-{\rm sign}_{ij}
{\rm sign}_{jk}
{\rm sign}_{ki}
\non\\
\label{idG}
\ear
the expression in braces is transformed into

\be
4{k_1\cdot k_2\over T}(G_{B13}+G_{B23}-G_{B12})
\label{rewritebraces}
\ee
This is precisely the same expression which appears also in the
exponential factor in (\ref{divAAA5}). After the usual rescaling
to the unit circle, and performance of the trivial $T$ - integral,
we find therefore a complete cancellation between the Feynman
numerator and denominator polynomials
\footnote{The on-shell conditions are not necessary
for this cancellation to occur \cite{dimcsc}. Moreover
the same mechanism was found to work also for the
calculation of the chiral anomaly in $D=8$
\cite{fred}.}.
Thus without further integration we obtain
already the desired result for the anomalous divergence,

\be
k_3^{\rho}
\langle A^{\mu}A^{\nu}A_5^{\rho}\rangle
=
{8\over {(4\pi)}^2}
\varepsilon^{\mu\nu\kappa\lambda}k_1^{\kappa}k_2^{\lambda}
\label{PCAC}
\ee

\section{Constant External Fields}

\subsection{QED in a Constant External Field}

Let us extend this formalism to the practically
important case of QED in a constant external field.
In field theory the Dirac equation in a constant field
can be solved exactly, so that one can absorb the
field already at the level of the Feynman rules
by a redefinition of the electron propagator
\cite{geheniau,tsai74I,bakast}.
Quite analogously in the worldline formalism we
can absorb a constant field into the worldline
Green's functions \cite{mckshe,cadhdu,shaisultanov,rss}.

Let us thus assume that we have, in addition to the
background field $A^{\mu}(x)$ which serves as a generator of the
external photons,
an additional field $\bar A^{\mu}(x)$
with constant field strength tensor
$\bar F_{\mu\nu}$. 
Using Fock--Schwinger gauge
centered at $x_0$ we may
take $\bar A^{\mu}(x)$ to be of the form

\begin{equation}
\bar A_{\mu}(x) = 
{1\over 2}y_{\nu}\bar F_{\nu\mu}\\
\label{fockschwinger}
\end{equation}

\noindent
The constant field contribution to the 
worldline Lagrangian may then be written
as

\begin{equation}
\Delta{\cal L} = {1\over 2}iey_{\mu}\bar F_{\mu\nu}
\dot y_{\nu} - ie\psi_{\mu}\bar F_{\mu\nu}\psi_{\nu}\\
\label{DeltaLkomp}
\end{equation}
\noindent
Since it is quadratic in the worldline
fields it need not be considered as part
of the interaction Lagrangian; we can absorb it
into the free worldline propagators. 
The details are given in \cite{rss};
the result is that the worldline Green's functions
$G_B,G_F$ should, in a constant external
electromagnetic field,
be replaced by (deleting the ``bar''
and abbreviating $z=eFT$)

\bear
{{\cal G}_B}(\tau_1,\tau_2) &=&
{T\over 2z^2}
\biggl({z\over{{\rm sin}(z)}}
\,{\rm e}^{-iz\dot G_{B12}}
+ iz\dot G_{B12} - 1\biggr)
\non\\
{\cal G}_{F}(\tau_1,\tau_2) &=&
G_{F12} {{\rm e}^{-iz\dot G_{B12}}\over {\rm cos}(z)}
\non\\
\label{calGBGF}
\ear\no
These expressions should be understood as power
series in the field strength matrix $F$.
They are, in general, nontrivial Lorentz matrices,
so that the Wick contraction rules eq.(\ref{wickrules})
have to be replaced by

\begin{eqnarray}
\langle y^{\mu}(\tau_1)y^{\nu}(\tau_2)\rangle
&=&
-{\cal G}_B^{\mu\nu}(\tau_1,\tau_2)\nonumber\\
\langle\psi^{\mu}(\tau_1)\psi^{\nu}(\tau_2)\rangle
&=&
\frac{1}{2}{\cal G}_F^{\mu\nu}(\tau_1,\tau_2)\nonumber\\
\label{exfieldGreen's}
\end{eqnarray}
\noindent
We also need the
generalization of $\dot G_B$,
which is

\begin{eqnarray}
\dot{\cal G}_B(\tau_1,\tau_2)
&=&
{i\over z}\biggl({z\over{{\rm sin}(z)}}
{\rm e}^{-iz\dot G_{B12}}-1\biggr)
\label{derivcalGB}
\end{eqnarray}
\noindent
The substitution
rule eq.(~\ref{subrule}) continues to hold. 
However,
in contrast to 
the vacuum worldline correlators $\dot G_B,G_F$
their generalizations
$\dot {\cal G}_B,{\cal G}_F$ have
non-vanishing coincidence limits,

\begin{eqnarray}
\dot {\cal G}_B(\tau,\tau) &=& i{\rm cot}(z)
-{i\over z}\\
{\cal G}_F(\tau,\tau) &=& -i\,{\rm tan}(z)\\
\label{coincidence2}\nonumber
\end{eqnarray}
\noindent
As a consequence the rule must be
extended to include one-cycles,

\begin{equation}
\dot{\cal G}_B(\tau_i,\tau_i)\rightarrow
\dot{\cal G}_B(\tau_i,\tau_i)
-{\cal G}_F(\tau_i,\tau_i)
\label{onecycle}
\end{equation}
\no
This is almost all one needs to know for computing one-loop
photon scattering amplitudes, or the corresponding
effective action, in a constant overall background field. 
The only further information required at the one--loop
level is the change in the free path integral determinants
due to the external field. 
This change is [18]

\bear
\!\!\!\!
{(4\pi T)}^{-{D\over 2}}
&\rightarrow&
{(4\pi T)}^{-{D\over 2}}
{\rm det}^{-{1\over 2}}
\biggl[{\sin(z)\over {z}}
\biggr] \quad
\label{scaldetext}\\
\!\!\!\!
{(4\pi T)}^{-{D\over 2}}
&\rightarrow&
{(4\pi T)}^{-{D\over 2}}
{\rm det}^{-{1\over 2}}
\biggl[{\tan(z)\over {z }}
\biggr] \quad
\label{spindetext}
\ear\no
for Scalar and Spinor QED, respectively.
Those determinants describe the vacuum amplitude in
a constant field, and therefore are
just the
proper-time integrands of the well-known 
Euler-Hei\-sen\-berg-Schwinger formulas
\cite{eulhei,schwinger}.

With this machinery set up it is then easy to derive
the following generalization of the master formula
for $N$ - photon scattering, eq.(\ref{scalarqedmaster}),
to the constant field case,

\bear
\Gamma[k_1,\varepsilon_1;\ldots;k_N,\varepsilon_N]
\!
&=&
\!
{(-ie)}^N
{\dps\int_{0}^{\infty}}{dT\over T}
{[4\pi T]}^{-{D\over 2}}
\non\\&&\hspace{-80pt}
\times {\rm e}^{-m^2T}
{\rm det}^{-{1\over 2}}
\biggl[{{\rm sin}(z)\over z}\biggr]
\times
\prod_{i=1}^N \int_0^T 
d\tau_i
\non\\&&\hspace{-80pt}
\times\exp\biggl\lbrace\sum_{i,j=1}^N 
\bigl\lbrack \half k_i\cdot {\cal G}_{Bij}\cdot  k_j
-i\varepsilon_i\cdot\dot{\cal G}_{Bij}\cdot k_j
\non\\&& \hspace{-35pt}
+\half
\varepsilon_i\cdot\ddot {\cal G}_{Bij}\cdot\varepsilon_j
\bigr\rbrack\biggr\rbrace
\mid_{\rm multi-linear}\quad
\non\\
\label{spinorqedmasterF}
\ear
Expanding out this formula for $N=2$, and performing
the partial integration procedure, we obtain the following
representation for the 
(dimensionally regularised) Scalar QED vacuum polarisation
tensor in a constant field,

\bear
\Pi^{\mu\nu}_{\rm scal}[k]
&=&
-{e^2\over {[4\pi]}^{D\over 2}}
{\dps\int_{0}^{\infty}}{dT\over T}
{T}^{2-{D\over 2}}
e^{-m^2T}
\non\\ &&\hspace{-20pt}\times
{\rm det}^{-\half}\biggl[{\sin(z)\over {z}}
\biggr] 
\int_0^1 du_1
\,\,I^{\mu\nu}_{\rm scal}
\,\e^{-Tk\cdot\Phi_{12}\cdot k}
\non
\ear

\bear
I^{\mu\nu}_{\rm scal}
&=&
\biggl\lbrace
\Bigl\lbrack
\dot{\cal S}^{\mu\nu}_{B12}\dot{\cal S}^{\kappa\lambda}_{B12}
- \dot{\cal S}^{\mu\lambda}_{B12}\dot{\cal S}^{\nu\kappa}_{B12}
\Bigr\rbrack
\non\\ &&\hspace{-25pt}
+
\Bigl(
\dot{\cal A}_{B12}-\dot{\cal A}_{B11}
\Bigr)^{\mu\lambda}
\Bigl(
\dot{\cal A}_{B12}-\dot{\cal A}_{B22}
\Bigr)^{\nu\kappa}
\biggr\rbrace
k^{\kappa}k^{\lambda}
\non\\
\label{vpFscal}
\ear\no
We have decomposed ${\cal G}_B$ 
and ${\cal G}_F$ into
their parts symmetric (``S'') and antisymmetric (``A'')
in the Lorentz indices,
${\cal G}_{B,F} = {\cal S}_{B,F} + {\cal A}_{B,F}$.

The application of the substitution rule to this
representation gives us the corresponding quantity
for Spinor QED,

\bear
\Pi^{\mu\nu}_{\rm spin}[k]
&=&
2{e^2\over {[4\pi]}^{D\over 2}}
{\dps\int_{0}^{\infty}}{dT\over T}
{T}^{2-{D\over 2}}
e^{-m^2T}
\non\\ && \hspace{-20pt}\times
{\rm det}^{-\half}\biggl[{\tan(z)\over {z}}\biggr]
\int_0^1 du_1
\,\,I^{\mu\nu}_{\rm spin}
\,\e^{-Tk\cdot\Phi_{12}\cdot k}
\non
\ear\no

\bear
I^{\mu\nu}_{\rm spin}
&=&
\biggl\lbrace
\Bigl\lbrack
\dot{\cal S}^{\mu\nu}_{B12}\dot{\cal S}^{\kappa\lambda}_{B12}
- \dot{\cal S}^{\mu\lambda}_{B12}\dot{\cal S}^{\nu\kappa}_{B12}
\Bigr\rbrack
\non\\&&
-
\Bigl\lbrack
{\cal S}^{\mu\nu}_{F12}{\cal S}^{\kappa\lambda}_{F12}
- {\cal S}^{\mu\lambda}_{F12}{\cal S}^{\nu\kappa}_{F12}
\Bigr\rbrack
\non\\
&&
+
\Bigl(
\dot{\cal A}_{B12}-\dot{\cal A}_{B11}+{\cal A}_{F11}
\Bigr)^{\mu\lambda}
\non\\&&\times
\Bigl(
\dot{\cal A}_{B12}-\dot{\cal A}_{B22}+{\cal A}_{F22}
\Bigr)^{\nu\kappa}
\non\\&&
-{\cal A}^{\mu\lambda}_{F12}
{\cal A}^{\nu\kappa}_{F12}
\biggr\rbrace
k^{\kappa}k^{\lambda}
\non\\
\label{vpFspin}
\ear\no
From (5.3),(5.5) we have

\bear
\dot{\cal A}_{Bij}&=&
i\Bigl(
{\cos(z\dot G_{Bij})\over\sin(z)}
-{1\over z}
\Bigr)\non\\
\dot{\cal A}_{Bii}&=&
i\Bigl(
\cot(z)
-{1\over z}
\Bigr)\non\\
\dot{\cal S}_{Bij}&=&
{\sin(z\dot G_{Bij})
\over
\sin(z)}\non\\
{\cal A}_{Fij}&=&
-iG_{Fij}{\sin(z\dot G_{Bij})\over\cos(z)}\non\\
{\cal A}_{Fii}&=& -i\tan (z)
\non\\
{\cal S}_{Fij}&=&
G_{Fij}
{\cos(z\dot G_{Bij})\over\cos(z)}
\non\\
\Phi_{ij}&=&
{1\over 2iz}\Bigl(\dot{\cal A}_{Bij}-\dot{\cal A}_{Bii}\Bigr)
\non\\
&=& {\cos(z\dot G_{Bij})-\cos(z)\over 2z\sin(z)}
\non\\
\label{defcoeffvp}
\ear
In writing eqs.(\ref{vpFscal}),(\ref{vpFspin}) we have already
re\-sca\-led to the unit circle and put
$u_2=0$.
Both $\Pi_{\rm scal}^{\mu\nu}$ \cite{bafrsh}
and $\Pi_{\rm spin}^{\mu\nu}$ \cite{batsha,artimovich}
were obtained before in field theory,
though with significantly more computational effort.
A similar gain in efficiency was found for the $N=3$ case
\cite{adlsch}. The three-point
calculation is of relevance for the
process of photon splitting in strong magnetic fields
\cite{adler71,stoneham,bamish}, another process of interest to
astrophysicists \cite{barhar}.

\subsection{The Vector-Axialvector Polarisation Tensor
in a Constant Field}

The above treatment of the constant external field
carries over to the full vector-axialvector case without
difficulty. We have used it for obtaining the
following representation for
the vector -- axialvector two-point amplitude
in a constant electromagnetic field
\cite{ioasch},

\bear
\langle
A^{\mu}(k_1)A^{\nu}_5(k_2)
\rangle_F
&=&
{(2\pi)}^4
\delta(k_1+k_2)
\Pi_5^{\mu\nu}
(k)
\non\\
\Pi_5^{\mu\nu}(k)
&=&
\half
{ee_5\over {[4\pi]}^2}
\Tintm
\non\\&& 
\times
\int_0^1d u_1
\,\,
I^{\mu\nu}_5
\,\e^{-Tk \Phi_{12} k} \nonumber
\non
\ear

\bear
I^{\mu\nu}_5 &=&
4ieT\biggl\lbrace
\tilde F^{\mu\nu}k{\cal U}_{12}k
-(\dot{\cal S}_{B12}\tilde F)^{\mu\nu}k\dot{\cal S}_{B12}k
\non\\
&&
+\Bigl[
(\tilde F k)^{\mu}({\cal U}_{12}k)^{\nu}
+ (\mu\leftrightarrow\nu )
\Bigr]
\non\\
&&
-\Bigl[
(\tilde F \dot{\cal S}_{B12}k)^{\nu}(\dot{\cal S}_{B12}k)^{\mu}
+ (\mu\leftrightarrow\nu )
\Bigr]
\biggr\rbrace
\non\\
&&\hspace{-8pt}
+(eT)^2\FFdual
\biggl\lbrace
-\dot{\cal A}_{B12}^{\mu\nu}k{\cal U}_{12}k
+(\dot{\cal S}_{B12}\dot{\cal A}_{B22})^{\mu\nu}k\dot{\cal S}_{B12}k
\non\\&&
-\Bigl[
(\dot{\cal A}_{B12}k)^{\mu}({\cal U}_{12}k)^{\nu}
+ (\mu\leftrightarrow\nu )
\Bigr]
\non\\
&&
+\Bigl[
(\dot{\cal S}_{B12}k)^{\mu}(\dot{\cal A}_{B22}\dot{\cal S}_{B12}k)^{\nu}
+ (\mu\leftrightarrow\nu )
\Bigr]
\biggr\rbrace
\label{defJmunu}
\ear\no
Here we introduced one more coefficient
function,

\bear
{\cal U}_{ij} &=& \dot{\cal S}_{Bij}^2 - 
(\dot{\cal A}_{Bij}
-\dot{\cal A}_{Bii})\bigl(\dot{\cal A}_{Bij}+{i\over z}\bigr)
\non\\
  &=&  {1-\cos(z\dot G_{Bij})\cos(z)
   \over \sin^2(z)} \non\\
\label{defU}
\ear\no
($\tilde F_{\mu\nu}=\half\varepsilon_{\mu\nu\alpha\beta}
F^{\alpha\beta}$).
As far as is known to the author this calculation was done
in field theory only for the special case of a pure
magnetic field \cite{demida,ioaraf}. In addition to being more
general the calculation presented here has also the
advantage of being manifestly (vector) gauge invariant.
In \cite{demida,ioaraf} gauge invariance 
was reached only after
performing a certain integration by parts, and appealing to
the absence of the chiral anomaly to be able to drop the
boundary term.

As was explained in the introduction, in the Fermi limit 
the calculation of the $\gamma\to\nu\bar\nu$ and
$\nu\to\nu\gamma$ processes in a constant magnetic field
essentially reduces to
a determination of $\Pi_{\rm spin}$ and $\Pi_5$. 
The further analysis shows
that for the plasmon
decay $\Pi$ dominates over $\Pi_5$, while for the
Cherenkov process it is the other way round
\cite{ioaraf}.

\section{Conclusions}

We have presented a recently
developed approach to calculations involving a fermion
loop and both vector and axialvector couplings \cite{mcksch,dimcsc}.
This formalism extends the second order formulation of QED
to the mixed vector-axialvector case. 
It also provides the corresponding generalization of the
QED worldline formalism, leading to a generalization of the
QED Bern-Kosower master formula \cite{dimcsc}. 
As in the QED case this formalism allows one to incorporate 
constant external electromagnetic
fields in a very economic way.

We discussed the properties of the formalism with regard to
the treatment of $\gamma_5$ in the dimensional continuation.
There is no breaking of the chiral symmetry for parity-even
loops. For parity-odd loops the usual expression is
obtained for the chiral anomaly,
however the anomalous divergences are
unambiguously confined to the axialvector legs.

We demonstrated that this technique provides
an easy and elegant way for calculating,
in a general constant electromagnetic field,
both the
ordinary photon polarisation tensor 
and the vector-axialvector polarisation tensor.
The relevance of these quantities to standard model
photon-neutrino processes has been discussed in
some detail.

While our representation here concentrated on the calculcation
of amplitudes, the formalism applies as well to the
calculation of the effective action eq.(\ref{defL}) itself.
The first few terms in the heat-kernel expansion of this
effective action were already presented in \cite{mcksch}. 
Work in this direction is in progress.

\end{document}